\documentclass[prl,twocolumn,amssymb]{revtex4}
\usepackage{graphicx}

\begin{document}

\title{Superfluidity in Three-species Mixture of Fermi Gases across Feshbach Resonances}
\author{Hui Zhai}
\affiliation {Department of Physics, the Ohio-State University, Columbus, 43210, OH}
\date{\today}
\begin{abstract}
In this letter a generalization of the BEC-BCS crossover theory to a multicomponent superfluid is presented by studying a three-species mixture of Fermi gas across two Feshbach resonances. At the BEC side of resonances, two kinds of molecules are stable which gives rise to a two-component Bose condensate. This two-component superfluid state can be experimentally identified from the radio-frequency spectroscopy, density profile and short noise measurements. As approaching the BCS side of resonances, the superfluidity will break down at some point and yield a first-order quantum phase transition to normal state, due to the mismatch of three Fermi surfaces. Phase separation instability will occur around the critical regime.

\end{abstract}
\maketitle 

{\it Introduction.}
During recent years, one major progress in quantum gases is achieving the crossover from a Bose condensation of molecules to a fermionic BCS superfluid in two-species fermionic gases across Feshbach resonance. Meanwhile, the multi-component Bose gases also attract considerable interests, mainly because of the effects related to internal phase coherence. Here we arise an interesting question that how the crossover process happens if the molecular BEC is a multi-component one, and it will be intriguing to achieve a fermionic multi-component superfluid which should be more attractive than its bosonic counterpart.
 
\begin{figure}[bp]
\begin{center}
\includegraphics[width=9.0cm]
{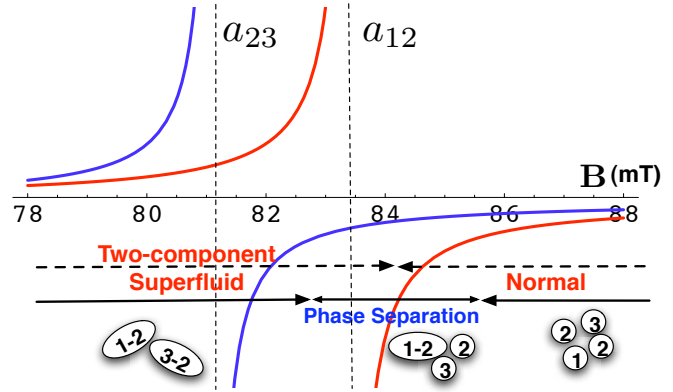}
\caption{An illustration of the scattering lengths $a_{12}$ and $a_{23}$ as a function of magnetic field and the phase diagram for the three-species mixture with $N_2=N_1+N_3$. The relation between scattering lengths and magnetic field is obtained from the measurement of Ref.\cite{Liscattering}. In this range of magnetic field $a_{13}$ is much smaller other two, and is not shown here. Restricted in homogeneous states, a first order quantum phase transition form the two-component superfluid state to the normal state is found above the first resonance. Incorporating imhomogeneous states, the phase separated state will take over around the critical magnetic field. The locations of phase boundary will depend on the density and the concentration.\label{phasediagram}}
\end{center}
\end{figure}

Lucky, this question is not only an academic one, but also close relates to current experiments on lithium and potassium gases. Recently an accurate measurement of the scattering lengths between the lowest three hyperfine spin states of lithium, denoted as $|1\rangle$, $|2\rangle$ and $|3\rangle$ respectively, finds that there is a Feshbach resonances between $|2\rangle-|3\rangle$ at $81.1{\text mT}$ followed by another resonance between $|1\rangle-|2\rangle$ at $83.4{\text mT}$, as shown in Fig.\ref{phasediagram}\cite{Liscattering}. Similar feature has also been found between three hyperfine spin states of potassium, where two resonances are located at $20.21{\text mT}$ and $22.42{\text mT}$ respectively\cite{Kscattering}. Note that both the $|1\rangle-|2\rangle$ molecule and the $|2\rangle-|3\rangle$ molecule are stable at the BEC sides of two resonances\cite{phrase}, we consider the case that the numbers of atoms in different species satisfy $N_2=N_1+N_3$, therefore we have a two-component molecular condensate deeply in the BEC side. The main purpose of this letter is to present a mean-field theory to describe two-component BEC-BCS crossover using lithium gas as an example\cite{Leggett}. 

On the other hand, it is also interesting to look at the system from the BCS side. Note that the normal state has three different Fermi surfaces, how the system copes with gaining paring energy and Fermi surface mismatch is a long-standing problem for both condensed matter and high energy physics. Recent experiments on two-species mixture with population imbalance\cite{experiments} have revealed many interesting phenomena such as superfluid-normal transition and phase separation, and also caused a considerable amount of theoretical interests\cite{theory}. Another point of this letter is to demonstrate that similar effects will also occur in the three-species mixture as it approaches the BCS side. A global phase diagram constructed by the mean-field theory is shown in Fig.\ref{phasediagram}.

Before proceeding, we would like to remark on the stability of the system against atom loss, which is the major concern of experiments. There are three sources of atom loss: atom-molecule inelastic collision, molecule-molecule inelastic collision and the Efimov states. In order to reduce the atom-molecule inelastic collision, one can initially prepare the system deeply in the BEC side where all the atoms are bound and the number of unbound atoms is very few, and then adiabatically tune to the resonance regime. The inelastic collision between molecules could be largely suppressed by the Pauli exclusion principle as all molecules contain species $|2\rangle$. As for the Efimov states, it is unlikely that it will significantly affect the three-body loss rate in this case.(see detailed discussion in footnote.\cite{Efimov} and Ref.\cite{Braaten}). Although the stability of this system remains to be seen in future experiments, qualitative speaking it is very promising that it can have a reasonable long life time for experimental study, and we hope the theoretical discussed below will motivate more research in this system.

{\it Mean-field Theory of Two-component Superfluid.} Following the idea of crossover theory\cite{BEC-BCSmf}, we write down a BCS-like wave function which contains pairing between both $|1\rangle-|2\rangle$ and $|2\rangle-|3\rangle$ \cite{wavefunction}
\begin{equation}
|\Psi\rangle=\prod_{{\bf k}}\left(u_{{\bf k}}+v_{{\bf k}}\psi^\dag_{{\bf k}2}\psi^\dag_{{\bf -k}1}+w_{{\bf k}}\psi^\dag_{{\bf k}2}\psi^\dag_{{\bf -k}3}\right)|0\rangle.\label{wavefunction}
\end{equation}
Note that the interaction between $|1\rangle-|3\rangle$ is much weak in the range of magnetic field from $78{\text mT}$ to $88{\text mT}$, comparing to those between $|1\rangle-|2\rangle$ and $|2\rangle-|3\rangle$\cite{Liscattering,Kscattering}, we neglect the pairing between $|1\rangle-|3\rangle$ because both of them try to pair with $|2\rangle$ first for energetic consideration, and also because the critical temperature to achieve $|1\rangle-|3\rangle$ pairing is too low to reach in current experiments. 

Here we would like to emphasize the symmetry and order parameters of this state. First, there are two independent pairing order parameters, $\Delta_1=\langle\psi^\dag_{2}\psi^\dag_{1}\rangle$ and $\Delta_2=\langle\psi^\dag_{2}\psi^\dag_{3}\rangle$. Furthermore, as both species $|1\rangle$ and $|3\rangle$ pair with species $|2\rangle$, the operator $\psi^\dag_3\psi_1$, which in fact converts a $|2\rangle-|1\rangle$ pair into a $|2\rangle-|3\rangle$ pair, acts as a Josephson tunneling between two components, and the wave function Eq.(\ref{wavefunction}) automatically gives another order parameter $\eta=\langle\psi^\dag_3\psi_1\rangle$ which is related to the relative phase between two components.

The Hamiltonian under consideration is written as
\begin{equation}
{\mathcal H}=\sum\limits_{{\bf k}\sigma}\epsilon_{{\bf k}\sigma}\psi^\dag_{{\bf k}\sigma}\psi_{{\bf k}\sigma}+\sum\limits_{{\bf k},{\bf k^\prime},i}g_{2i}\psi^\dag_{{\bf k}2}\psi^\dag_{{\bf -k}i}\psi_{{\bf -k^\prime}i}\psi_{{\bf k^\prime}2},\label{Hamiltonian}
\end{equation}
where $\epsilon_{{\bf k}\sigma}=\hbar^2 k^2/(2m)-\mu_{\sigma}$, $\sigma=1,2,3$ and $i=1,3$. As in the two-species mixture, $g_{2i}$ is related to the scattering lengthes via
$1/g_{2i}=m/(4\pi \hbar^2 a_{2i})-\sum_{{\bf k}}m/(\hbar^2 k^2)$. The interaction between $|1\rangle-|3\rangle$, which is neglected here, can be turned on as perturbation in a more detailed study elsewhere, and it will not affect the qualitative features discussed here. 

With the Hamiltonian Eq.(\ref{Hamiltonian}) the free-energy for the quantum state of Eq.(\ref{wavefunction}) is given by
\begin{eqnarray}
\mathcal{F}=&&\sum\limits_{{\bf k}}\left(\xi_{{\bf k}a}|v_{{\bf k}}|^2+\xi_{{\bf k}b}|w_{{\bf k}}|^2\right)\nonumber \\ &&+\sum\limits_{{\bf k}{\bf k^\prime}}g_{21}u_{{\bf k}}v^*_{{\bf k}}u^*_{{\bf k^\prime}}v_{{\bf k^\prime}}+g_{23}u_{{\bf k}}w^*_{{\bf k}}u^*_{{\bf k^\prime}}w_{{\bf k^\prime}},
\end{eqnarray}
where $\xi_{{\bf k}a}=\epsilon_{{\bf k}1}+\epsilon_{{\bf k}2}$ and $\xi_{{\bf k}b}=\epsilon_{{\bf k}2}+\epsilon_{{\bf k}3}$. The constraint $|u_{{\bf k}}|^2+|v_{{\bf k}}|^2+|w_{{\bf k}}|^2=1$ can be imposed by a Langrange multiplier $\sum_{{\bf k}}\lambda_{{\bf k}}(|u_{{\bf k}}|^2+|v_{{\bf k}}|^2+|w_{{\bf k}}|^2-1)$. Minimizing the free-energy with respect to $u_{{\bf k}}$, $v_{{\bf k}}$ and $w_{{\bf k}}$ gives $\partial{\mathcal{F}}/\partial u_{{\bf k}}=\partial{\mathcal{F}}/\partial v_{{\bf k}}=\partial{\mathcal{F}}/\partial w_{{\bf k}}=0$, which are
\begin{equation}
\left(\begin{array}{ccc}\lambda_{{\bf k}} & \Delta_1 & \Delta_2 \\\Delta_1^* & \lambda_{{\bf k}}-\xi_{{\bf k}a} & 0 \\\Delta_2^* & 0 & \lambda_{{\bf k}}-\xi_{{\bf k}b}\end{array}\right)\left(\begin{array}{c}u_{{\bf k}} \\v_{{\bf k}} \\w_{{\bf k}}\end{array}\right)=0,\label{minimization}
\end{equation}
where $\Delta_1=-g_{21}\sum_{{\bf k}}u_{{\bf k}}v^*_{{\bf k}}$ and $\Delta_2=-g_{23}\sum_{{\bf k}}u_{{\bf k}}w^*_{{\bf k}}$.
Thus $\lambda_{{\bf k}}$ satisfies the equation 
\begin{equation}
\lambda_{{\bf k}}^3-A_{{\bf k}}\lambda^2_{{\bf k}}+B_{{\bf k}}\lambda_{{\bf k}}+C_{{\bf k}}=0,\label{cubic}
\end{equation}
where $A_{{\bf k}}=\xi_{{\bf k}a}+\xi_{{\bf k}b}$, $B_{{\bf k}}=\xi_{{\bf k}a}\xi_{{\bf k}b}-|\Delta_1|^2-|\Delta_2|^2$ and $C_{{\bf k}}=|\Delta_1|^2\xi_{{\bf k}b}+|\Delta_2|^2\xi_{{\bf k}a}$. The lowest solution of Eq.(\ref{cubic}) is 
\begin{equation}
\lambda_{{\bf k}}=\frac{1}{3}\{A_{{\bf k}}-2\sqrt{A_{{\bf k}}^2-3B_{{\bf k}}}\cos[(\pi-\theta_{{\bf k}})/3]\},
\end{equation}
where $\theta_{{\bf k}}=\arctan[3\sqrt{3 K_{{\bf k}}}/(2A^3_{{\bf k}}-9A_{{\bf k}}B_{{\bf k}}-27C_{{\bf k}})]$ and $K_{{\bf k}}=A_{{\bf k}}^2B_{{\bf k}}^2-4B_{{\bf k}}^3+4A^3_{{\bf k}}C_{{\bf k}}-18A_{{\bf k}}B_{{\bf k}}C_{{\bf k}}+27C^2_{{\bf k}}$\cite{cubic}.

\begin{figure}[tbp]
\begin{center}
\includegraphics[width=9.0cm]
{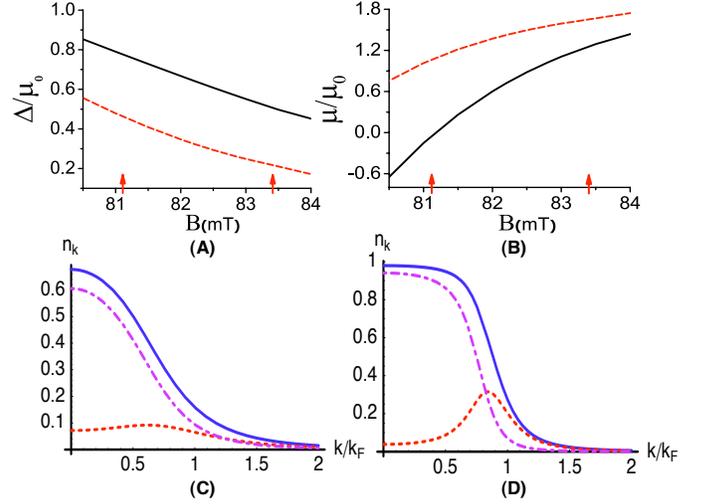}
\caption{The solution to the mean-field equations. (A-B): As the change of magnetic field,  the evolvement of two pairing order parameters $\Delta_1$(solid line) and $\Delta_2$(dashed line) (Fig.A), and the chemical potential $\mu_a=\mu_1+\mu_2$(solid line) and $\mu_b=\mu_2+\mu_3$(dashed line) (Fig.B). (C-D): The momentum distributions for $|1\rangle$(dashed line), $|2\rangle$(solid line) and $|3\rangle$(dashed-dotted line) at $80.5{\text mT}$(Fig.C) and $84{\text mT}$(Fig.D). Here we have chosen  $n_{2}=1.14\times 10^{11}\text{ cm}^{-3}$ and $N_{2}:N_1:N_3=1:0.5:0.5$. The two arrows in (A) and (B) indicate the locations of Feshbach resonances. $k_{{\text F}}$ is the Fermi momentum for the majority species $|2\rangle$, and $\mu_0=\hbar^2 k^2_{{\text F}}/(2m)$. \label{BEC-BCS}}
\end{center}
\end{figure}

For the wave function satisfying Eq.(\ref{minimization}), the free-energy is given by $\mathcal{F}=\sum_{{\bf k}}\lambda_{{\bf k}}-|\Delta_1|^2/g_{21}-|\Delta_2|^2/g_{23}$. It can be verified that the long wave length behavior of $\lambda_{{\bf k}}$ is precisely $-(|\Delta_1|^2+|\Delta_2|^2)m/(\hbar^2 k^2)$, therefore the divergency in the summation over $\lambda_{{\bf k}}$ can be exactly canceled out by the renormalization terms in $g_{21}$ and $g_{23}$. 
Furthermore, $\partial{\mathcal{F}}/\partial \Delta^*_j=0$ ($j=1,2$) yield two coupled self-consistency equations. Differential Eq.(\ref{cubic}) with respect to $\Delta^*_1$, one finds that $\partial\lambda_{{\bf k}}/\partial \Delta^*_1$ satisfies
\begin{equation}
(3\lambda^2_{{\bf k}}-2A_{{\bf k}}\lambda_{{\bf k}}+B_{{\bf k}})\frac{\partial \lambda_{{\bf k}}}{\partial \Delta^*_1}=\Delta_1(\lambda_{{\bf k}}-\xi_{{\bf k}b}).
\end{equation}
Thus, $\partial {\mathcal{F}}/\partial \Delta^*_1$ gives
\begin{equation}
\frac{m}{4\pi\hbar^2 a_{12}}=\sum\limits_{{\bf k}}\left(\frac{\lambda_{{\bf k}}-\xi_{{\bf k}b}}{3\lambda^2_{{\bf k}}-2A_{{\bf k}}\lambda_{{\bf k}}+B_{{\bf k}}}+\frac{1}{\hbar^2 k^2/m}\right),
\end{equation} 
and similarly we have
\begin{equation}
\frac{m}{4\pi \hbar^2 a_{23}}=\sum\limits_{{\bf k}}\left(\frac{\lambda_{{\bf k}}-\xi_{{\bf k}a}}{3\lambda^2_{{\bf k}}-2A_{{\bf k}}\lambda_{{\bf k}}+B_{{\bf k}}}+\frac{1}{\hbar^2 k^2/m}\right).
\end{equation}

Apart from increasing of gaps, another feature of the BEC-BCS crossover is the shift of chemical potentials, which is determined with the help of number equations. From Eq.(\ref{minimization}) we can get $u_{{\bf k}}/v_{{\bf k}}=(\xi_{{\bf k}a}-\lambda_{{\bf k}})/\Delta^*_1$ and $w_{{\bf k}}/v_{{\bf k}}=[\lambda_{{\bf k}}(\lambda_{{\bf k}}-\xi_{{\bf k}a})-|\Delta_1|^2]/(\Delta^*_1\Delta_2)$. Therefore, two number equations are given by $N_{1}=\sum_{{\bf k}}|v_{{\bf k}}|^2=\sum_{{\bf k}}(1/(1+|u_{{\bf k}}/v_{{\bf k}}|^2+|w_{{\bf k}}/v_{{\bf k}}|^2)$ and $N_{3}=\sum_{{\bf k}}|w_{{\bf k}}|^2=\sum_{{\bf k}}(|w_{{\bf k}}/v_{{\bf k}}|^2/(1+|u_{{\bf k}}/v_{{\bf k}}|^2+|w_{{\bf k}}/v_{{\bf k}}|^2)$ respectively.

We solve these four coupled equations numerically, and the results are shown in Fig.\ref{BEC-BCS}, which exhibit the common features of BEC-BCS crossover. From Fig.\ref{BEC-BCS}(A-B), one can see that the gaps increase and the chemical potentials decrease as the magnetic field changes from the BCS side to BEC side. Fig.\ref{BEC-BCS}(C-D) displays the momentum distributions for three different species.
One can see that the momentum distributions become much broader at the BEC side comparing to those at the BCS side, which indicates that the real space sizes of two kinds of Cooper pairs are much smaller. This solution also allows one to calculate the energy of this superfluid state $E_{\text {SF}}$.

{\it Quantum Phase Transition for Homogeneous System.} As we have mentioned, in the normal state the Fermi surface of species $|2\rangle$ is larger than those of species $|1\rangle$ and $|3\rangle$ for the concentration $N_2=N_1+N_3$, it will cost a lot of kinetic energy of two minority species to form pairs. Thus one expects that there is a quantum phase transition form this two-component superfluid state to the normal state as approaching BCS side.

\begin{figure}[tbp]
\begin{center}
\includegraphics[width=9.0cm]
{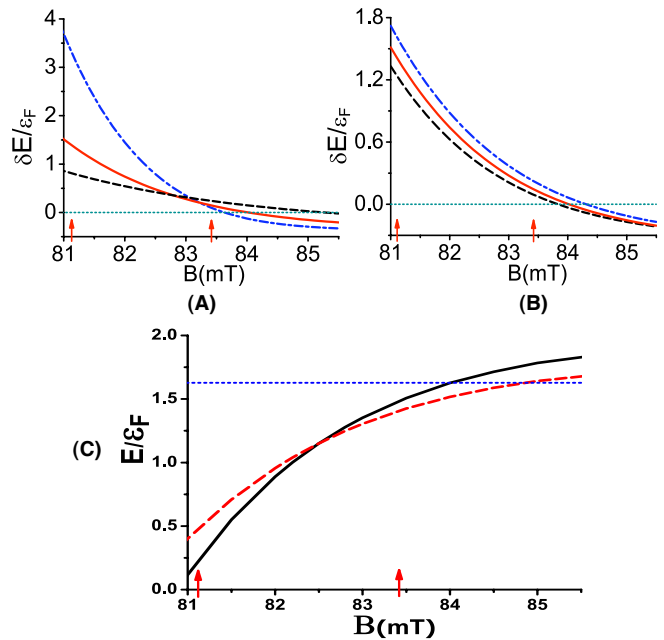}
\caption{(A-B) $\delta E$ as a function of magnetic field. (A): $N_2:N_1:N_3=1:0.5:0.5$ and $k_{{\text F}}a_{0}=1\times 10^{-4}$(solid line), $k_{{\text F}}a_{0}=0.5\times 10^{-4}$(dashed line), $k_{{\text F}}a_{0}=2\times 10^{-4}$(dashed-dotted line). (B) $k_{{\text F}}a_{0}=1\times 10^{-4}$ and $N_{2}:N_1:N_3=1:0.5:0.5$(solid line), $N_2:N_1:N_3=1:0.4:0.6$(dashed line), $N_2:N_1:N_3=1:0.6:0.4$(dashed-dotted line). (C) Comparison of the energy between the phase separated state(dashed line), the two-component superfluid state(solid line) and the normal state(dotted line). $k_{\text{F}}a_0=1\times 10^{-4}$ and $N_2:N_1:N_3=1:0.5:0.5$. For all figures the unit of energy density is $\mathcal{E}_{{\text F}}=\hbar^2 k^5_{{\text F}}/(20\pi^2 m)$, $k_{{\text F}}$ is the Fermi momentum of the major species $|2\rangle$ in its normal state, and $a_{0}=0.0529177{\text {nm}}$. Two arrows indicate the locations of Feshbach resonances.\label{phasetransition}}
\end{center}
\end{figure}

At mean-field level, we simply take the normal state as non-interacting normal state, and its energy $E_{{\text N}}$ is the total kinetic energy of three species. In Fig.\ref{phasetransition}(A-B) we plot the energy difference $\delta E=E_{\text{N}}-E_{\text {SF}}$ for different densities and concentrations. A quantum phase transition occurs when $\delta E$ becomes negative, i.e. the normal state has lower energy. The critical point is usually located around the first Feshbach resonance, and its exact location depends on density and concentration. We find that the critical point will be pushed toward the BCS side either when the density of atoms increases or when the radio $N_1:N_3$ increases\cite{recover}. We also notice that one can not connect the quantum state of the form Eq.(\ref{wavefunction}) to a normal state by continuous varying the parameters, therefore this transition should be a first order quantum phase transition.

{\it Phase Separation Instability.} There are possible phase separation instability in both normal state and the two-component superfluid state. Approaching the first resonance from the normal state, $a_{12}$ becomes larger while $a_{23}$ is still not larger, the energy may be lowered if the normal state separates into a phase separated state constructed as follows: in one region it is $\prod_{{\bf k}}(u_{{\bf k}}+v_{{\bf k}}\psi^\dag_{{\bf k}2}\psi^\dag_{{\bf -k}1})\prod_{{\bf p=0}}^{p_{\text{F}}}\psi^\dag_{{\bf p}3}|0\rangle$ and in another region it is $\prod_{{\bf k}=0}^{k_{\text{F}}}\psi^\dag_{{\bf k}2}\prod_{{\bf p}=0}^{p_{\text{F}}}\psi^\dag_{{\bf p}3}|0\rangle$. The energy of this state is shown in Fig.\ref{phasetransition}(C) for a typical density and concentration. We find that around the critical magnetic field the energy of phase separated state is lower than both two homogeneous states.

In the superfluid phase $\kappa=\partial^2 \mathcal{F}/(\partial |\Delta_1|^2\partial |\Delta_2|^2)$ is always positive, indicating the interaction to be repulsive. Hence, one need to consider whether the repulsion is so strong that lead the two-component condensate spatially separating into two condensates, with $|2\rangle-|1\rangle$ molecules staying in one side and $|2\rangle-|3\rangle$ molecules staying in another side. Here we also calculate the energy of this type phase separated state, and find that the energy is very close to, but usually slightly higher than, the energy of the homogenous state in all range of magnetic field. 

Combining the discussion on the homogenous states and the phase separation instability, we draw the conclusion on the phase diagram illustrated in Fig.\ref{phasediagram}. We remark that the effects missing in the mean-field theory, including the interaction energy of normal state and quantum fluctuations in superfluid phase, will give quantitative corrections to this phase diagram. However, the qualitative feature, which is guaranteed by the physical understanding at two ends, should holds.

{\it Experimental Signatures of the Two-component Superfluid.} At the end of this paper, we point out some experimental signatures to reveal the unique features of the two-component supefluid state.

To detect the coherence between two components, one can look at the radio-frequency spectroscopy, which can be described by $H_{\text{rf}}=\psi^\dag_{3}\psi_{1}+\text{h.c.}$ and has already been widely used in fermion experiments\cite{rf}. Applying r-f field in this superfluid state will induce a time-perodic oscillation of the atom number in both species $|1\rangle$ and $|3\rangle$, which can be a direct evidence of phase coherence between two components.

A direct signal to distinguish this state from the normal state is an {\it in situ} measurement of the density profile. In the superfluid state the density profiles of different species have to satisfy $n_{2}({\bf r})=n_{1}({\bf r})+n_{3}({\bf r})$ everywhere, and the Thomas-Fermi radii are the same for all species. In contrast, in the normal state the Thomas-Fermi radius of the majority species is larger than those of two minority species, and $n_{2}({\bf r})$ does not equal to $n_{1}({\bf r})+n_{3}({\bf r})$ everywhere. For non-interacting normal state, at the center of the trap, $n_{2}/(n_1+n_3)=N_2^{2/3}/(N_1^{2/3}+N_3^{2/3})<1$.

Analyzing the noise correlation in time-of-flight image allows one to measure the second-order correlation function $\mathcal{G}_{\alpha\beta}({\bf k},{\bf k^\prime})=\langle \hat{n}_{\alpha{\bf k}}\hat{n}_{\beta{\bf k^\prime}}\rangle-\langle\hat{n}_{\alpha{\bf k}}\rangle\langle\hat{n}_{\beta{\bf k^\prime}}\rangle$\cite{noise}. In this state $\mathcal{G}_{21}({\bf k},{\bf k^\prime})=|u_{{\bf k}}v_{{\bf k}}|^2\delta({\bf k}+{\bf k^\prime})$ and $\mathcal{G}_{23}({\bf k},{\bf k^\prime})=|u_{{\bf k}}w_{{\bf k}}|^2\delta({\bf k}+{\bf k^\prime})$, which shows  strong pairing correlation between opposite momentums. Here, one remarkable feature is the effective exclusive correlation between $|1\rangle$ and $|3\rangle$, and it shows up in $\mathcal{G}_{13}({\bf k},{\bf k^\prime})=-|v_{{\bf k}}w_{{\bf k}}|^2\delta({\bf k}-{\bf k^\prime})$ as a dip at equal momentum. This arises from the fact  that the atom in the quantum state $|2,{\bf -k}\rangle$ can only pair with either $|1,{\bf k}\rangle$ or $|3,{\bf k}\rangle$, thus the quantum state with momentum ${\bf k}$ can not be occupied by both $|1\rangle$ and $|3\rangle$.

The author acknowledge Tin-Lun Ho for insightful discussion and helpful comments on the manuscript, Zuo-Zi Chen for numerical assistance, Cheng Chin for helpful discussion on the scattering properties of lithium and Eric Braaten for a helpful conversation on Efimov states. This work is supported by NSF Grant DMR-0426149 awarded to T.L.Ho.


\begin{thebibliography}{99}
\bibitem{phrase}
Following the terminology used in the discussion of two-species mixture, we call the high field regime as the BCS side and the low field regime as the BEC side.
\bibitem{recover}
In the limit $N_3\rightarrow 0$, the superfluid state will always has a lower energy than the normal state, which recovers the familiar two-species mixture case.
\end{thebibliography}
\end{document}